\documentclass[9pt,twocolumn,twoside]{opticajnl}

\journal{ol} 

\setboolean{shortarticle}{true}

\usepackage{lineno}
\usepackage{braket}
\usepackage{float}

\title{Optical Memory for Arbitrary Perfect Poincar\'{e} States in an Atomic Ensemble}

\author[1,2,4]{Lei Zeng}
\author[1,2,3,4]{Ying-Hao Ye}
\author[1,2,3,*]{Ming-Xin Dong}
\author[1,2]{Wei-Hang Zhang}
\author[1,2]{En-Ze Li}
\author[3,*]{Da-Chuang Li}
\author[1,2,*]{Dong-Sheng Ding}
\author[1,2]{Bao-Sen Shi}

\affil[1]{Key Laboratory of Quantum Information, University of Science and Technology of China, Hefei, Anhui 230026, China.}
\affil[2]{Synergetic Innovation Center of Quantum Information and Quantum Physics, University of Science and Technology of China, Hefei, Anhui 230026, China.}
\affil[3]{Institute for Quantum Control and Quantum Information and School of Physics and Materials Engineering, Hefei Normal University, Hefei, Anhui 230601, China.}
\affil[4]{These authors contributed equally to this work.}

\affil[*]{Corresponding author: dongmx@ustc.edu.cn}
\affil[*]{Corresponding author: dachuangli@ustc.edu.cn}
\affil[*]{Corresponding author: dds@ustc.edu.cn}

\begin{abstract}
Inherent spin angular momentum (SAM) and orbital angular momentum (OAM) which manifest as polarization and spatial degrees of freedom (DOF) of photons, hold a promise of large capability for applications in classical and quantum information processing. To enable these photonic spin and orbital dynamic properties strongly coupled with each other, Poincar\'{e} states have been proposed and offer advantages in data multiplexing, information encryption, precision metrology, and quantum memory. However, since the transverse size of Laguerre Gaussian beams strongly depends on their topological charge numbers $\left| l \right|$, it is difficult to store asymmetric Poincar\'{e} states due to the significantly different light-matter interaction for distinct spatial modes. Here, we experimentally realize the storage of perfect Poincar\'{e} states with arbitrary OAM quanta using the perfect optical vortex, in which 121 arbitrarily-selected perfect Poincar\'{e} states have been stored with high fidelity. The reported work has great prospects in optical communication and quantum networks for dramatically increased encoding flexibility of information.
\end{abstract}

\setboolean{displaycopyright}{true}

\begin{document}

\maketitle

As robust information carriers, photons have many degrees of freedoms (DOFs), such as polarization \cite{ding2015raman,xu2013long}, spatial-mode \cite{ding2013single,nicolas2014quantum}, temporal mode \cite{heller2020cold} and path (k-vector) \cite{parniak2017wavevector,pu2017experimental}, which provides a dramatically increased encoding potential. Optical memories encoded with different photonic DOFs have attracted a wide range of attention in various information processing \cite{lvovsky2009optical}. Poincar\'{e} beams that simultaneously couple spin angular momentum (SAM) and orbital angular momentum (OAM) of photons have azimuthally-dependent polarization distribution across their cross-section \cite{berry2004,dennis2002,freund2002,arora2019}. They have been widely applied in many fields \cite{rosales2018}, such as optical trapping \cite{kawauchi2007}, optical communications \cite{milione2015,milione2015using}, laser material processing \cite{hamazaki2010} and optical encryption\cite{fang2020}. As the asymmetric characteristic of SAM and OAM for Poincar\'{e} states, the storage of Poincar\'{e} beams manifests the ability of the memories to simultaneously store two DOFs of photons, which is important for a practical memory. 
\begin{figure*}[ht]
	\centering
	\includegraphics[width=0.9 \linewidth]{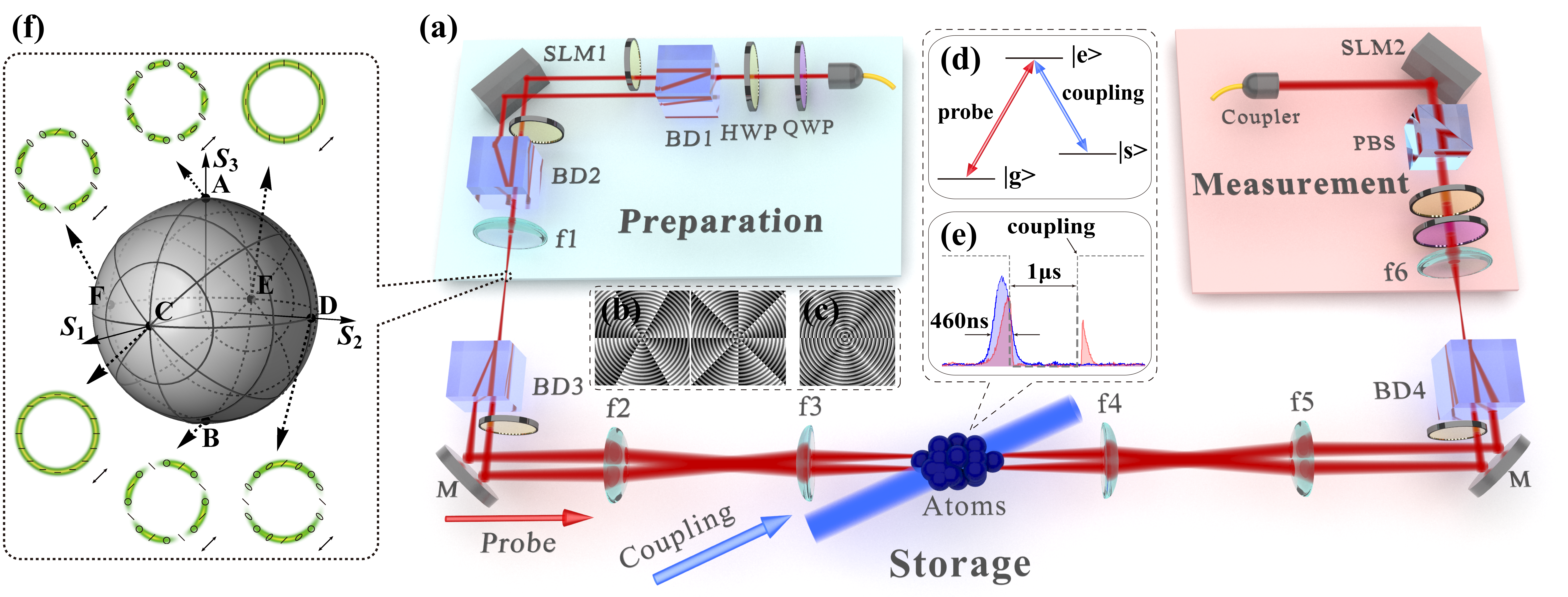}
	\caption{Experimental principle. (a) Setup. The focal lengths of lenses $f_1$, $f_2$, $f_3$, $f_4$, $f_5$ and $f_6$ are 75, 500, 300, 300, 500 and 75 mm, respectively.  PBS: polarizing beam splitter;  SLM: spatial light modulator; QWP: quarter-wave plate; HWP: half-wave plate; BD: beam displacer. (b),(c) Examples of the hologram on the SLM1 and SLM2. (d) Associated energy levels for EIT-based optical storage. $\ket{g}$,$\ket{s}$ and $\ket{e}$ correspond to the $^{85}$Rb atomic hyperfine levels $\ket{5S_{1/2},F=2}$, $\ket{5S_{1/2},F=3}$ and $\ket{5P_{1/2},F=3}$ respectively. (e) Time sequence of the optical storage. After a programmable storage time, the signal retrieved from the memory is sent to the measurement part. (f) The hybrid-order Poincar\'{e} Sphere (HyOPS) representation of various PPBs shown in the (H, V) basis. The polarization distributions are shown above the annular intensity profiles of PPBs which are transformed into distinct patterns using a linear polarizer depicted by the black double arrow. The two poles (C and E) are represent $\ket{H}\ket{POV_{i,L_1}}$ and $\ket{V}\ket{POV_{i,L_2}}$ $(i=1,2,3,4)$. A,B,D and F represent the states $\ket{\Psi_i}$ expressed in Eq.(\ref{eq2}).}
	\label{fig:1}
\end{figure*}

High-fidelity memories for OAM superposition states and conventional Poincar\'{e} States have been demonstrated in atomic systems recently \cite{ding2013single,nicolas2014quantum,parigi2015,ye2019,Ye2022}, however, those works only exploited the symmetric OAM superposition states and ignore the asymmetric contribution from different OAM states. Since the radius of a conventional Laguerre-Gaussian (LG) beam is proportional to $\sqrt{\left|l\right|+1}$ \cite{ding2014toward,ding2015quantum}, LG modes with  topological charge number $\lvert l\rvert$ interact with different 
regions of the atomic ensemble and thus experience different optical depths (ODs) of the ensemble. Due to the strong correlation between the storage efficiency and effective atomic OD \cite{hsiao2018}, memory for conventional Poincar\'{e} states with asymmetric OAM values suffers from this imbalanced efficiency. For instance, a photonic state $\ket{\Psi}_{\text{in}}=1/\sqrt{2}(\ket{L_1}+\ket{L_2})$ evolves to $\ket{\Psi}_{\text{re}}=1/\sqrt{\eta_1^2+\eta_2^2}\left(\eta_1\ket{L_1}+\eta_2\ket{L_2}\right)$ after being stored and retrieved and the corresponding fidelity is then reduced to $1/2+\kappa/(1+\kappa^2)$, here $\eta_1$, $\eta_2$ are storage efficiencies for $\ket{L_1}, \ket{L_2}$ and $\kappa=\eta_2/\eta_1$.
Therefore, these previous works are limited by the opposite topological charge numbers, which restrict the scenarios of Poincar\'{e} beams' application. The perfect optical vortex (POV) that generated from the Fourier transform of a Bessel Gaussian (BG) mode \cite{vaity2015,wang2017,li2019} helps us circumvent this obstacle because of its topological charge number-independent ring diameters \cite{ostrovsky2013}. Perfect Poincaré beams (PPBs) have been experimentally generated by using various optical components such as axicons, spatial light modulators (SLM), q-plates, lenses, and the micro-patterned liquid-crystal devices \cite{xu2018}.
After replacing the conventional LG modes with corresponding POV modes, one can achieve nearly identical efficiencies for the modes with different OAM quanta and then realize a faithful storage of perfect Poincar\'{e} beams (PPBs) with asymmetric OAM quanta.

Here, we demonstrate a coherent optical memory for perfect Poincar\'{e} states based on electromagnetically induced transparency (EIT) protocol \cite{chaneliere2005,eisaman2005,radwell2015spatially,hsiao2018} by constructing a spatial-mode-independent light-matter interface and realize the storage of different 121 perfect Poincar\'{e} states with an arbitrary choice of OAM quanta. Our work shows a dramatically increased encoding flexibility and is promising for classical information processing and quantum communication.

The setup can be divided into three parts: state preparation, optical storage, and projective measurement (see Fig.~\ref{fig:1}). The first part consists of two beam displacers (BDs), a lens $f_1$, and a spatial light modulator (SLM, HOLOEYE: LETO). The lens $f_1$ acting as a Fourier transformer to transform the BG states imprinted by the SLM to corresponding perfect Poincar\'{e} states. BD3 and a half-wave plate (HWP) convert PPBs into two beams with the same polarization. We ensure balanced storage efficiency for individual beams by adjusting the mirror. The lenses $f_2$ and $f_3$ constitute a telescope system that image the perfect Poincar\'{e} states to the center of the ensemble. The lens $f_5$, together with the lens $f_6$, also form a $4f$ imaging system to restore the probe beams to their original size for subsequent projective measurements. The storage is carried out in an ensemble of cold $^{85}$Rb trapped in a two-dimensional magneto-optical trap (MOT). The signal and control fields are both circularly polarized ($\sigma^+$), and the expanded coupling field has a radius of 4 mm to cover the probe field, and it intersects with the probe beams at the center of the ensemble with an angle of $5^{\circ}$. The measurement part is composed of a lens $f_6$, a half-wave plate (HWP), a quarter-wave plate (QWP), a polarizing beam splitter (PBS), and an SLM that is placed at the back focal plane of the lens $f_6$. The probe light is finally coupled into a single-mode fiber (SMF) for detection.

\begin{figure}[htbp]
	\centering
	\includegraphics[width=0.9\linewidth]{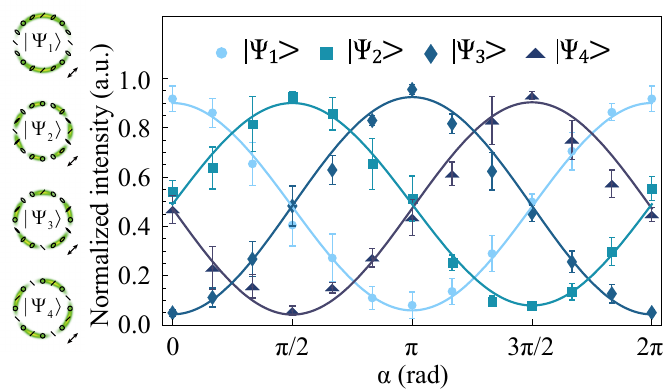}
	\caption{Interference curves for $\ket{\Psi_1}$,$\ket{\Psi_2}$,$\ket{\Psi_3}$ and $\ket{\Psi_4}$ represent the measured intensity for different values of the relative angle $\alpha$ when the signal is projected into states $\frac{1}{\sqrt{2}}(\ket{H}+\ket{V})$ after storage. The error bars represent $\pm$ s.d.}
	\label{fig:2}
\end{figure}

\begin{figure*}[htbp]
	\centering
	\includegraphics[width=\linewidth]{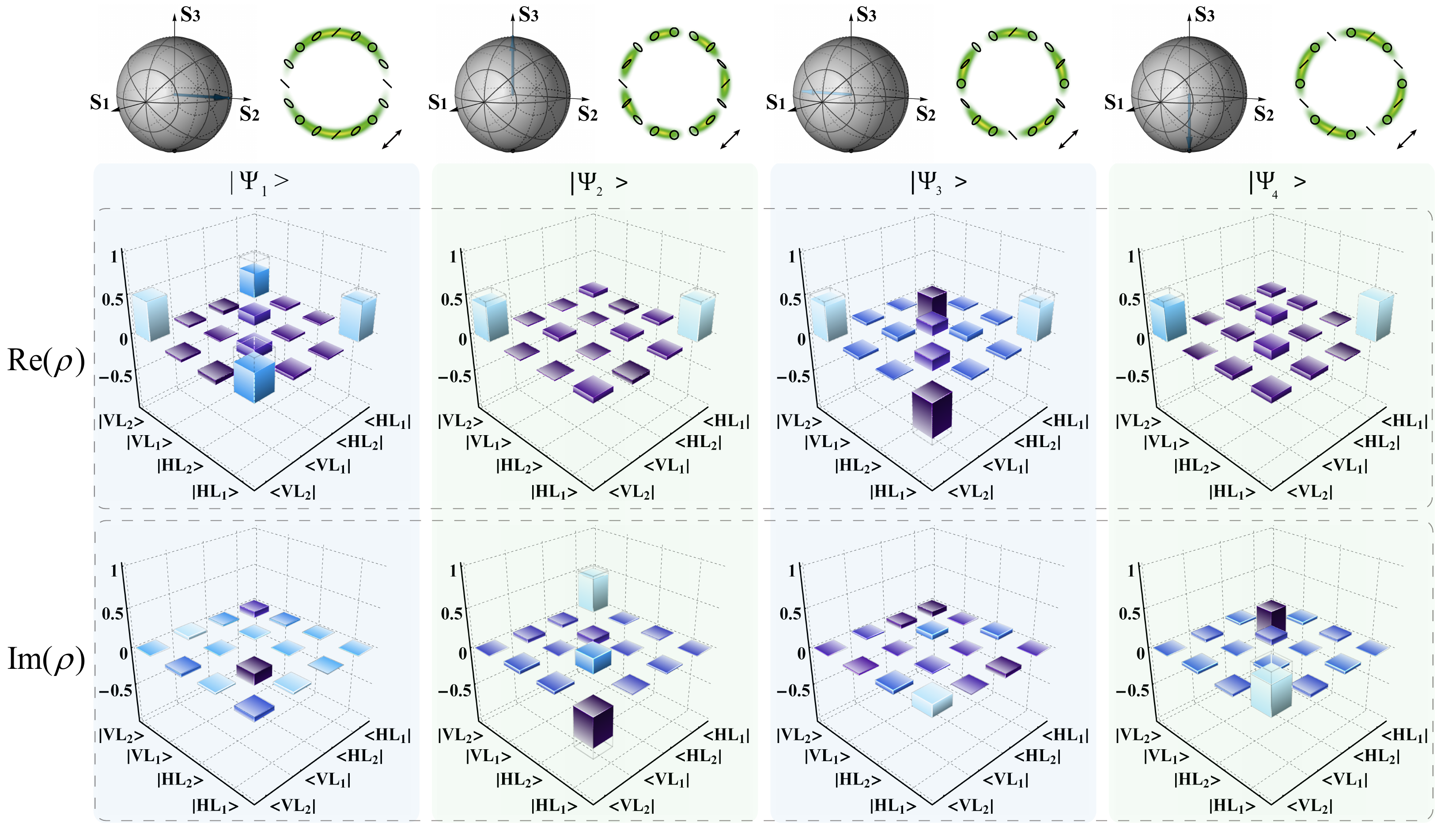}
	\caption{The density matrices of retrieved states for $\ket{\Psi_1}$,$\ket{\Psi_2}$,$\ket{\Psi_3}$ and $\ket{\Psi_4}$. The first row displays, for each state, its location in the HyOPS, and the associated polarization profiles. The second and the third rows correspond to the real and imaginary parts of the reconstructed density matrix $\rho$, respectively, and the black line boxes in the figure correspond to the ideal optical states.}
	\label{fig:3}
\end{figure*}

The POV can be generated from the Fourier transform of a higher-order Bessel beam \cite{ostrovsky2013}, for the situation of large $r_r$ and small $\omega_0$, the complex field amplitude is expressed by

\begin{equation}
E^l_{POV}(r,\varphi)=i^{l-1}\frac{2f}{k\omega_0^2}\text{exp}(il\varphi)\text{exp}\left(-\frac{(r-r_r)^2}{\omega_0^2}\right)
\label{eq1}
\end{equation}
where $\omega_0$ is the waist of the corresponding Gaussian mode at the focal plane of the Fourier and $k=2\pi/\lambda$ is the wave number. Eq. (\ref{eq1}) represents the field amplitude of a POV with radius equal to $r_r(=k_rf/k)$, $l$ is the topological charge number and $k_r$ is the radial wave vector. The radius of the POV is independent of $l$, this property is crucial to the realization of mode-independent light-matter interaction and leads to the same storage efficiency against different OAM quanta. The achieved efficiency remains nearly unchanged in a wide range of $l$ and decreases slightly when $\left| l \right|$ exceeds 10. Therefore the range of $l$ is chosen from -5 to 5 in our experiments since the storage efficiency ($14.3\%\pm 0.4\%$) barely changes within this range.

We characterize the retrieved states by performing projective measurements. The polarization base is $(\ket{H}+\ket{V})/\sqrt{2}$, and the OAM base is set to be $\ket{POV_{i,L_1}}+e^{i\alpha}\ket{POV_{i,L_2}}$, where $\alpha$ is the relative phase between $\ket{POV_{i,L_1}}$ and $\ket{POV_{i,L_2}}$. The state of signal light is tailored to be perfect Poincar\'{e} state $\ket{\Psi_i}=\ket{H}\ket{POV_{i,L_1}}+e^{i\theta}\ket{V}\ket{POV_{i,L_2}}$ initially, after being projected onto the above basis, the intensity of the retrieved signal that collected by the SMF is proportional to $1+\text{cos}(\theta-\alpha)$. By tilting the 
BD3, the values of $\theta$ are set as $0,\pi/2,\pi$ and $3\pi/2$, then we can achieve the four target perfect Poincar\'{e} states:
\begin{equation}
\begin{aligned}
&\ket{\Psi_1}=\ket{H}\ket{POV_{1,L_1=1}}+\ket{V}\ket{POV_{1,L_2=3}},\\
&\ket{\Psi_2}=\ket{H}\ket{POV_{2,L_1=-3}}+i\ket{V}\ket{POV_{2,L_2=4}},\\
&\ket{\Psi_3}=\ket{H}\ket{POV_{3,L_1=0}}-\ket{V}\ket{POV_{3,L_2=-5}},\\
&\ket{\Psi_4}=\ket{H}\ket{POV_{4,L_1=2}}-i\ket{V}\ket{POV_{4,L_2=-2}}.
\end{aligned}
\label{eq2}
\end{equation}

Figure \ref{fig:2} shows the measured interference curve as a function of $\alpha$ after storage. The average interference visibility values of $\ket{\Psi_1}$, $\ket{\Psi_2}$, $\ket{\Psi_3}$ and $\ket{\Psi_4}$ are $0.838\pm0.050$, $0.844\pm0.034$, $0.900\pm0.033$ and $0.881\pm0.034$, respectively. A bias magnetic field of 1 G guides the magnetically-induced evolution of OAM modes of two separated components stored in atoms \cite{ye2021synchronized}. The experimental results show that the polarization and phase are preserved during storage.

The storage fidelity, which is defined as $[\text{Tr}(\sqrt{\sqrt{\rho}\rho_{0}\sqrt{\rho}})]^2$, is an important criterion to characterize the similarity between the retrieved signal ($\rho$) and the original signal ($\rho_{0}$). The chosen polarization bases for reconstructing the density matrix are $\ket{H}$,$\ket{V}$,$\ket{H}+i\ket{V}$ and $\ket{H}+\ket{V}$, the chosen bases in OAM degree of freedom are  $\ket{L_1}$,$\ket{L_2}$,$\ket{L_1}+i\ket{L_2}$ and $\ket{L_1}+\ket{L_2}$. Figure \ref{fig:3} shows the results of the reconstructed density matrix for these four optical states. The measured results of the fidelity of these four states are $81.1\%\pm4.7\%$, $84.4\%\pm4.5\%$, $82.5\%\pm4.3\%$, and $86.7\%\pm3.4\%$, respectively.

\begin{figure}[htbp]
	\centering
	\includegraphics[width=\linewidth]{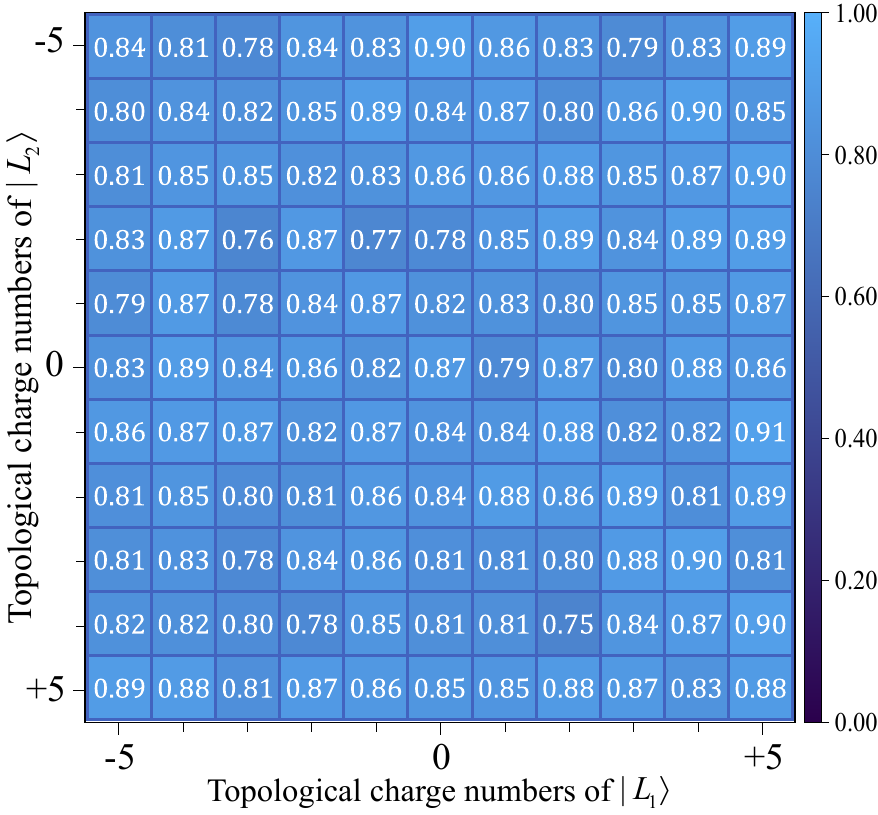}
	\caption{The estimated fidelity. The numbers marked in the rows and columns represent the OAM topological charge numbers of $\ket{L_1}$ and $\ket{L_2}$ ranging from -5 to 5 respectively.}
	\label{fig:4}
\end{figure}

Then we evaluate the memory of perfect Poincar\'{e} states with arbitrary $L_1, L_2\in [-5,5]$ by estimating fidelities of 121 PPBs. For the OAM DOF, we set the projected polarization basis to be $\ket{H}+\ket{V}$, and the phase diagram corresponding to the highest (lowest) theoretical value of the projective measurement is loaded on SLM2. Similarly, we get the visibility for polarization DOF by selecting the OAM base $\ket{POV_{L_1}}+\ket{POV_{L_2}}$ and choosing the highest (lowest) polarization base. Then the average interference visibility $V$ can be achieved (for the states with $L_1=L_2$, we take the visibility for polarization DOF as its average interference visibility), so the fidelity can be quickly estimated as $F=(1+3V)/4$ \cite{arahira20121}. The achieved storage fidelities of these states are shown in Fig.~\ref{fig:4}, which indicate the capability of our memory to store Poincar\'{e} light composed of two arbitrary different OAM quanta.

We report a work on optical memory of perfect Poincar\'{e} states based on atomic ensemble, the results show that the system holds the ability of OAM quanta independent storage efficiency. Assuming the cold atom system can store $d$ orthogonal different OAM modes, and there are $2d-1$ orthogonal states with a symmetric value of topological charge numbers suitable for encoding in the case of the conventional Poincar\'{e} states, while for the perfect Poincar\'{e} states, the number of states appropriate for encoding increases to $d^2$. This significantly increases the flexibility of coding choices.

It is possible to implement optical storage for perfect Poincar\'{e} states to avoid using the complex dual-rail setup by employing alkali
atoms such as $^{87}$Rb with simpler Zeeman sublevels structure. With the help of higher requirements for frequency shifting and magnetic field manipulation, one may provide the potential in simplifying the system through lifting the Zeeman-degeneration of hyper-fine ground states\cite{xu2013long}.
The storage time of the system is 1.5 $\mu$s, which can be extended with the help of the techniques of magnetically-insensitive states preparation. \cite{zhao2009millisecond}.

In conclusion, we realize an optical memory for perfect Poincar\'{e} states in an atomic ensemble. By using the transverse-size-invariance of perfect Poincar\'{e} states, we construct a spatial-mode-independent light-matter interface, and realize the storage of 121 perfect Poincar\'{e} states with arbitrary OAM. Our work largely extends the encoding flexibility of information, which is promising for quantum communication and information processing. 

\begin{backmatter}
	\bmsection{Funding}
	This work was supported by National Key R\&D Program
	of China (Grants No. 2017YFA0304800), Anhui Initiative in Quantum Information Technologies (Grant
	No.~AHY020200), the National Natural Science
	Foundation of China (Grants No. U20A20218, No. 61722510, No. 11934013, No. 11604322, No. 12204461),
	and the Innovation Fund from CAS, the Youth Innovation Promotion Association
	of CAS under Grant No. 2018490, the Anhui Provincial Key Research and Development Project under Grant No. 2022b13020002, and
	the Anhui Provincial Candidates for academic and technical leaders Foundation under Grant No. 2019H208.

	\bmsection{Acknowledgment}
	We thank Prof. Wei Zhang for helpful discussions.

	\bmsection{Disclosures}
	The authors declare no competing interests.
	
	\bmsection{Data Availability Statement}
	Data underlying the results presented in this paper are not publicly available at this time but may be obtained from the authors upon reasonable request.
	
\end{backmatter}



\begin{thebibliography}{10}
\newcommand{\enquote}[1]{``#1''}

\bibitem{ding2015raman}
D.-S. Ding, W.~Zhang, Z.-Y. Zhou, S.~Shi, B.-S. Shi, and G.-C. Guo,
  {\protect\JournalTitle{Nature Photonics}} \textbf{9}, 332 (2015).

\bibitem{xu2013long}
Z.~Xu, Y.~Wu, L.~Tian, L.~Chen, Z.~Zhang, Z.~Yan, S.~Li, H.~Wang, C.~Xie, and
  K.~Peng, {\protect\JournalTitle{Physical Review Letters}} \textbf{111},
  240503 (2013).

\bibitem{ding2013single}
D.-S. Ding, Z.-Y. Zhou, B.-S. Shi, and G.-C. Guo, {\protect\JournalTitle{Nature
  Communications}} \textbf{4}, 1 (2013).

\bibitem{nicolas2014quantum}
A.~Nicolas, L.~Veissier, L.~Giner, E.~Giacobino, D.~Maxein, and J.~Laurat,
  {\protect\JournalTitle{Nature Photonics}} \textbf{8}, 234 (2014).

\bibitem{heller2020cold}
L.~Heller, P.~Farrera, G.~Heinze, and H.~de~Riedmatten,
  {\protect\JournalTitle{Physical Review Letters}} \textbf{124}, 210504 (2020).

\bibitem{parniak2017wavevector}
M.~Parniak, M.~D{\k{a}}browski, M.~Mazelanik, A.~Leszczy{\'n}ski, M.~Lipka, and
  W.~Wasilewski, {\protect\JournalTitle{Nature Communications}} \textbf{8}, 1
  (2017).

\bibitem{pu2017experimental}
Y.~Pu, N.~Jiang, W.~Chang, H.~Yang, C.~Li, and L.~Duan,
  {\protect\JournalTitle{Nature Communications}} \textbf{8}, 1 (2017).

\bibitem{lvovsky2009optical}
A.~I. Lvovsky, B.~C. Sanders, and W.~Tittel, {\protect\JournalTitle{Nature
  Photonics}} \textbf{3}, 706 (2009).

\bibitem{berry2004}
M.~Berry, {\protect\JournalTitle{Journal of Optics A: Pure and Applied Optics}}
  \textbf{6}, 475 (2004).

\bibitem{dennis2002}
M.~Dennis, {\protect\JournalTitle{Optics Communications}} \textbf{213}, 201
  (2002).

\bibitem{freund2002}
I.~Freund, {\protect\JournalTitle{Optics Communications}} \textbf{201}, 251
  (2002).

\bibitem{arora2019}
G.~Arora, P.~Senthilkumaran \emph{et~al.}, {\protect\JournalTitle{Optics
  Letters}} \textbf{44}, 5638 (2019).

\bibitem{rosales2018}
C.~Rosales-Guzm{\'a}n, B.~Ndagano, and A.~Forbes,
  {\protect\JournalTitle{Journal of Optics}} \textbf{20}, 123001 (2018).

\bibitem{kawauchi2007}
H.~Kawauchi, K.~Yonezawa, Y.~Kozawa, and S.~Sato, {\protect\JournalTitle{Optics
  Letters}} \textbf{32}, 1839 (2007).

\bibitem{milione2015}
G.~Milione, M.~P. Lavery, H.~Huang, Y.~Ren, G.~Xie, T.~A. Nguyen, E.~Karimi,
  L.~Marrucci, D.~A. Nolan, R.~R. Alfano \emph{et~al.},
  {\protect\JournalTitle{Optics Letters}} \textbf{40}, 1980 (2015).

\bibitem{milione2015using}
G.~Milione, T.~A. Nguyen, J.~Leach, D.~A. Nolan, and R.~R. Alfano,
  {\protect\JournalTitle{Optics Letters}} \textbf{40}, 4887 (2015).

\bibitem{hamazaki2010}
J.~Hamazaki, R.~Morita, K.~Chujo, Y.~Kobayashi, S.~Tanda, and T.~Omatsu,
  {\protect\JournalTitle{Optics Express}} \textbf{18}, 2144 (2010).

\bibitem{fang2020}
X.~Fang, H.~Ren, and M.~Gu, {\protect\JournalTitle{Nature Photonics}}
  \textbf{14}, 102 (2020).

\bibitem{parigi2015}
V.~Parigi, V.~D’Ambrosio, C.~Arnold, L.~Marrucci, F.~Sciarrino, and
  J.~Laurat, {\protect\JournalTitle{Nature Communications}} \textbf{6}, 1
  (2015).

\bibitem{ye2019}
Y.-H. Ye, M.-X. Dong, Y.-C. Yu, D.-S. Ding, and B.-S. Shi,
  {\protect\JournalTitle{Optics Letters}} \textbf{44}, 1528 (2019).

\bibitem{Ye2022}
Y.-H. Ye, L.~Zeng, M.-X. Dong, W.-H. Zhang, E.-Z. Li, D.-C. Li, G.-C. Guo,
  D.-S. Ding, and B.-S. Shi, {\protect\JournalTitle{Phys. Rev. Lett.}}
  \textbf{129}, 193601 (2022).

\bibitem{ding2014toward}
D.-S. Ding, W.~Zhang, Z.-Y. Zhou, S.~Shi, J.-s. Pan, G.-Y. Xiang, X.-S. Wang,
  Y.-K. Jiang, B.-S. Shi, and G.-C. Guo, {\protect\JournalTitle{Physical Review
  A}} \textbf{90}, 042301 (2014).

\bibitem{ding2015quantum}
D.-S. Ding, W.~Zhang, Z.-Y. Zhou, S.~Shi, G.-Y. Xiang, X.-S. Wang, Y.-K. Jiang,
  B.-S. Shi, and G.-C. Guo, {\protect\JournalTitle{Physical Review Letters}}
  \textbf{114}, 050502 (2015).

\bibitem{hsiao2018}
Y.-F. Hsiao, P.-J. Tsai, H.-S. Chen, S.-X. Lin, C.-C. Hung, C.-H. Lee, Y.-H.
  Chen, Y.-F. Chen, A.~Y. Ite, and Y.-C. Chen, {\protect\JournalTitle{Physical
  Review Letters}} \textbf{120}, 183602 (2018).

\bibitem{vaity2015}
P.~Vaity and L.~Rusch, {\protect\JournalTitle{Optics Letters}} \textbf{40}, 597
  (2015).

\bibitem{wang2017}
T.~Wang, S.~Fu, F.~He, and C.~Gao, {\protect\JournalTitle{Applied Optics}}
  \textbf{56}, 7567 (2017).

\bibitem{li2019}
D.~Li, S.~Feng, S.~Nie, C.~Chang, J.~Ma, and C.~Yuan,
  {\protect\JournalTitle{Journal of Applied Physics}} \textbf{125}, 073105
  (2019).

\bibitem{ostrovsky2013}
A.~S. Ostrovsky, C.~Rickenstorff-Parrao, and V.~Arriz{\'o}n,
  {\protect\JournalTitle{Optics Letters}} \textbf{38}, 534 (2013).

\bibitem{xu2018}
R.~Xu, P.~Chen, J.~Tang, W.~Duan, S.-J. Ge, L.-L. Ma, R.-X. Wu, W.~Hu, and
  Y.-Q. Lu, {\protect\JournalTitle{Physical Review Applied}} \textbf{10},
  034061 (2018).

\bibitem{chaneliere2005}
T.~Chaneli{\`e}re, D.~Matsukevich, S.~Jenkins, S.-Y. Lan, T.~Kennedy, and
  A.~Kuzmich, {\protect\JournalTitle{Nature}} \textbf{438}, 833 (2005).

\bibitem{eisaman2005}
M.~D. Eisaman, A.~Andr{\'e}, F.~Massou, M.~Fleischhauer, A.~S. Zibrov, and
  M.~D. Lukin, {\protect\JournalTitle{Nature}} \textbf{438}, 837 (2005).

\bibitem{radwell2015spatially}
N.~Radwell, T.~W. Clark, B.~Piccirillo, S.~M. Barnett, and S.~Franke-Arnold,
  {\protect\JournalTitle{Physical Review Letters}} \textbf{114}, 123603 (2015).

\bibitem{ye2021synchronized}
Y.-H. Ye, L.~Zeng, Y.-C. Yu, M.-X. Dong, E.-Z. Li, W.-H. Zhang, Z.-K. Liu,
  L.-H. Zhang, G.-C. Guo, D.-S. Ding \emph{et~al.},
  {\protect\JournalTitle{Physical Review A}} \textbf{103}, 053316 (2021).

\bibitem{arahira20121}
S.~Arahira, T.~Kishimoto, and H.~Murai, {\protect\JournalTitle{Optics Express}}
  \textbf{20}, 9862 (2012).

\bibitem{zhao2009millisecond}
B.~Zhao, Y.-A. Chen, X.-H. Bao, T.~Strassel, C.-S. Chuu, X.-M. Jin,
  J.~Schmiedmayer, Z.-S. Yuan, S.~Chen, and J.-W. Pan,
  {\protect\JournalTitle{Nature Physics}} \textbf{5}, 95 (2009).

\end{thebibliography}



\ifthenelse{\equal{\journalref}{aop}}{%
\section*{Author Biographies}
\begingroup
\setlength\intextsep{0pt}
\begin{minipage}[t][6.3cm][t]{1.0\textwidth} 
  \begin{wrapfigure}{L}{0.25\textwidth}
    \includegraphics[width=0.25\textwidth]{john_smith.eps}
  \end{wrapfigure}
  \noindent
  {\bfseries John Smith} received his BSc (Mathematics) in 2000 from The University of Maryland. His research interests include lasers and optics.
\end{minipage}
\begin{minipage}{1.0\textwidth}
  \begin{wrapfigure}{L}{0.25\textwidth}
    \includegraphics[width=0.25\textwidth]{alice_smith.eps}
  \end{wrapfigure}
  \noindent
  {\bfseries Alice Smith} also received her BSc (Mathematics) in 2000 from The University of Maryland. Her research interests also include lasers and optics.
\end{minipage}
\endgroup
}{}

\end{document}